\documentclass[aps,preprint]{revtex4}
\usepackage{amsmath}

\usepackage{amsfonts}

\input{tcilatex}

\begin{document}

\title[Waveform Selection Based Upon Post-Selection Criteria]{A Method for
Selecting Sensor Waveforms Based Upon Post-Selection Criteria for Remote
Sensing Applications}
\author{John E Gray}
\affiliation{Code Q-31, Electromagnetic and Sensor Systems Department, Naval Surface
Warfare Center Dahlgren, 18444 FRONTAGE ROAD SUITE 328, DAHLGREN VA
22448-5161}
\author{Allen D Parks}
\affiliation{Code Q-31, Electromagnetic and Sensor Systems Department, Naval Surface
Warfare Center Dahlgren, 18444 FRONTAGE ROAD SUITE 328, DAHLGREN VA
22448-5161}

\begin{abstract}
In previous work, we have argued that measurement using a radar can be
viewed as taking the expected value of an operator. The operator usually
represents some aspect of the characteristics of the object being tracked
(such as Doppler, distance, shape, polarization, etc.) that is measured by
the radar while the expectation is taken with respect to an optimal matched
filter design process based on the waveform broadcast by the radar and a
receiver which is optimized to a specific characteristic of the object being
tracked. With digital technology, it is possible to produce designer
waveforms both to broadcast and to mix with the return signal, so it is
possible to determine the maximum of the expectation of the operator by
proper choice of the received signal. We illustrate a method for selecting
the choice of the return signal to detect different "target operators" using
perturbation theory based on the Matched Filter Principle and illustrate it
with different operators and waveforms.
\end{abstract}

\maketitle
\tableofcontents

\section{Introduction}

In the seminal book "Probability and Information Theory with Applications to
Radar"\cite{Woodward1980}, Woodward introduced the ambiguity function as the
means to solve the measurement problem of radar. The measurement problem of
an active sensor is to design a waveform to be broadcast by a radar or
sonar, to maximize the receiver response to the signal which has interacted
with an object. The solution proposed by North\cite{North1943} during World
War II is the "matched filter", which correlates a known signal template
with what is received in a return signal to detect the presence or absence
of the template in the unknown received signal. This is exactly equivalent
to convolving the unknown signal with the complex conjugate of the
time-reversed version of the known signal template; this is called
cross-correlation. Therefore, as has been shown in many texts\cite%
{VanTrees2002a}, the matched filter is the optimal linear filter for
maximizing the signal to noise ratio (SNR) in the presence of additive noise.

In radar or sonar, a known signal is sent out and the reflected signal from
the object (which is a function of the distance to the object, the relative
speed of the object and the broadcast frequency of the radar), can be
examined at the radar receiver for the common elements of the out-going
signal in the return signal, which, when optimized is a multi-dimensional
matched filter or ambiguity function. The broadband form the return signal
is $s(\alpha t-\tau )$, where, $\tau=\frac{2R}{c}$, is the delay 
\begin{equation}
\alpha=\frac{c-v_{R}}{c+v_{R}}=\frac{1-\beta}{1+\beta}.
\end{equation}
Here $c$ is the speed of propagation and $v_{R}$ is the radial velocity of
the object.

There are two forms for the ambiguity function, the more general form is the
wideband (WB): where the return signal can be modeled as a delay in time of
the broadcast signal. The wideband ambiguity function, $\chi_{WB}$, which
has the return signal modeled as both a dilation and delay of the broadcast
signal 
\begin{equation}
\chi_{WB}(\omega,\tau)=\int_{-\infty}^{\infty}e^{-it\omega}\,s^{\ast
}(t)s_{N,WB}(t)\,dt,
\end{equation}
where $s^{\ast}(t)$ means complex conjugate of the broadcast signal. The
ambiguity function is used to design radar signals so that they have
desirable properties useful for various kinds of radars (Leavon is a current
up to date resource\cite{Levanon2004}). We propose a way to think about the
ambiguity function which is different than the way Woodward presented it.
This approach suggests the ambiguity function can be thought of as the
expectation value of an operator that is connected to the delay and dilation
properties associated with the Doppler effect \cite{Gray2010}. Thus, the
sensor measurement problem can be cast in a more abstract setting, which
treats interaction between the waveform and the target as an operator acting
on the waveform. This approach can be termed the operator approach and it
can be viewed as an abstraction of the quantum mechanical formalism applied
to a classical setting. This approach underlies the time-frequency approach
to signal processing that has been championed by Cohen\cite{Cohen1995}.
Using this approach, we examine the operator viewpoint for both single and
multi-dimensional operators acting on a signal by the interaction process.
In particular, we propose that the cross-ambiguity function for certain
operators can be used to amplify the return signals. We illustrate this for
several operators, show under what conditions this amplification can occur,
and discuss how the cross-amplification signal can be constructed given
knowledge of the interaction operator and the broadcast signal. The result
of this approach is to suggest a way for recasting problems in signal
processing when we have sufficient knowledge of the interaction of the
broadcast signal.

\section{Operator Approach}

The notation for the \textbf{inner product} of two signals, $r(t)$ and $s(t)$
that is used throughout the paper is 
\begin{equation}
\left\langle r(t),s(t)\right\rangle =\int_{-\infty}^{\infty}r^{\ast
}(t)\,s(t)\;dt,
\end{equation}
while the \textbf{Fourier transform, }$\mathcal{F}$\textbf{,} of a signal $%
s(t)$ is\cite{Papoulis1961} 
\begin{equation}
S(\omega)=\mathcal{\hat{F}}s(t)=\int_{-\infty}^{\infty}e^{-it\omega
}\,s(t)\;dt=\left\langle \,e^{it\omega},s(t)\right\rangle ,
\end{equation}
and the inverse Fourier transform, $\mathcal{\hat{F}}^{-1}$, is 
\[
s(t)=\mathcal{\hat{F}}^{-1}S(\omega)=\frac{1}{2\pi}\int_{-\infty}^{\infty
}e^{it\omega}\,S(\omega)d\omega=\left\langle \,e^{-it\omega},S(\omega
)\right\rangle . 
\]
A function of time which is translated by amount, $\tau,$ can be written as
(using the Taylor expansion of function $\hat{D}=\frac{d}{dt}$) 
\begin{equation}
s(t+\tau)=e^{\tau\hat{D}}s(t)=e^{i\left( -i\tau\frac{d}{dt}\right) }s(t)=e^{i%
\mathcal{\hat{W}}}s(t)
\end{equation}
The form of the narrow band (N) ambiguity function $\chi_{N}(\omega,\tau)$,
can be recast as 
\begin{equation}
\chi_{N}(\omega,\tau)=\int_{-\infty}^{\infty}e^{-it\omega}\,s^{\ast
}(t)s(t-\tau)\,dt=\left\langle s(t)e^{it\omega},e^{-i\tau\mathcal{\hat{W}}%
}s(t)\right\rangle .
\end{equation}
From the Doppler effect perspective, translation is the operation of the
frequency operator on the signal, where $\tau$ is the total distance a
signal travels to an object, is reflected, and then returns to the receiver.
The expected value associated with observable, $\hat{A},$ for a signal $%
s\left( t\right) $ is%
\begin{equation}
\left\langle \hat{A}\right\rangle =\int\hat{A}\,\left\vert s\left( t\right)
\right\vert ^{2}dt=\int s^{\ast}\left( t\right) \,\hat{A}s\left( t\right)
\,dt=\left\langle s\left( t\right) ,\hat{A}s\left( t\right) \right\rangle .
\end{equation}
Thus, the narrow band ambiguity function can be written using this
definition as%
\[
\chi_{N}(\omega,\tau)=\left\langle e^{-i\tau\mathcal{\hat{W}}}\right\rangle
_{s\left( t\right) }. 
\]
We can thus interpret $e^{\pm i\tau\mathcal{\hat{W}}}$ as a translation
operation acting on function $s(t)$ which moves the time $t\rightarrow
t\pm\tau$. This way of considering measurement in radar is a natural
continuation of the viewpoint that started with Gabour\cite{Gabor1953} and
extended by Woodward\cite{Woodward1980} and Vaidman\cite{Vakman1968} for
considering measurement in radar.

The time operator, $\mathcal{\hat{T}},$ is%
\begin{equation}
\mathcal{\hat{T}}=-\frac{1}{i}\frac{d}{d\omega},
\end{equation}
while the frequency operator is 
\begin{equation}
\mathcal{\hat{W}=}\frac{1}{i}\frac{d}{dt}.
\end{equation}
It is understood that these operators act on signals and that 
\begin{equation}
\mathcal{\hat{W}}^{n}s\left( t\right) =\left( \frac{1}{i}\frac{d}{dt}\right)
^{n}s\left( t\right) .
\end{equation}
A very useful calculation trick is based on a modification of Parceval's
theorem for an unnormalized signal: 
\begin{align}
E & =\left\langle s(t),s(t)\right\rangle =\left\langle 1\right\rangle
_{s\left( t\right) }  \nonumber \\
& =\frac{1}{2\pi}\left\langle \left\langle S(\omega^{\prime})e^{i\omega
^{\prime}t},1\right\rangle ,\left\langle S(\omega)e^{i\omega
t},1\right\rangle ,1\right\rangle  \nonumber \\
& =\left\langle \left\langle S(\omega^{\prime}),S(\omega)\delta\left(
\omega-\omega^{\prime}\right) \right\rangle \right\rangle  \nonumber \\
& =\left\langle S(\omega),S(\omega)\right\rangle =\left\langle
1\right\rangle _{S}.
\end{align}
Now it follows that the expected value of the frequency of a signal $S\left(
\omega\right) $ can be written as 
\begin{equation}
\left\langle \omega\right\rangle =\text{ }_{s(t^{\prime})}\left\langle 
\mathcal{\hat{W}}\right\rangle _{s(t)}.  \nonumber
\end{equation}
From this result, it follows that 
\begin{equation}
\left\langle \omega^{n}\right\rangle =\text{ }_{s(t^{\prime})}\left\langle 
\mathcal{\hat{W}}^{n}\right\rangle _{s(t)},
\end{equation}
which can be proved by induction. If $g\left( t\right) $ is an analytical
function, it follows that 
\begin{equation}
\left\langle g\left( \omega\right) \right\rangle =\text{ }%
_{s(t^{\prime})}\left\langle g\left( \mathcal{\hat{W}}\right) \right\rangle
_{s(t)}.
\end{equation}
Thus, to calculate the average frequency of a function, we do not have to
calculate the Fourier transform. Rather one simply calculates derivatives of
a function and then integrates.

The frequency translation operator has exactly the same effect: 
\begin{equation}
e^{i\theta\mathcal{\hat{T}}}S\left( \omega\right) =S\left( \omega
+\theta\right) .
\end{equation}
For a complex signal, $s\left( t\right) =A\left( t\right) e^{i\vartheta
\left( t\right) }$, 
\begin{equation}
e^{i\tau\mathcal{\hat{W}}}s(t)=\left( \vartheta^{\prime}\left( t\right) -i%
\frac{A^{\prime}\left( t\right) }{A\left( t\right) }\right) s(t)
\end{equation}
so 
\begin{equation}
\left\langle \omega\right\rangle _{S}=\left\langle \left( \vartheta^{\prime
}\left( t\right) +i\frac{A^{\prime}\left( t\right) }{A\left( t\right) }%
\right) A\left( t^{\prime}\right) ,A\left( t\right) \right\rangle
=\left\langle \vartheta^{\prime}\left( t\right) \right\rangle _{A\left(
t\right) }
\end{equation}
since the second term in the integral is a perfect differential. The average
frequency is the derivative of the phase, $\vartheta\left( t\right) $, over
the density over all time. Thus the phase at each time must be instantaneous
in some sense, i.e. $\omega_{i}\left( t\right) $, so we can make the
identification that $\omega_{i}\left( t\right) =\vartheta^{\prime}\left(
t\right) $. Similarly, we can show that 
\begin{equation}
\left\langle \omega^{2}\right\rangle _{S\left( \omega\right) }=\left\langle
\vartheta^{\prime2}\left( t\right) \right\rangle _{A\left( t\right)
}+\left\langle \frac{A^{\prime}\left( t\right) }{A\left( t\right) }%
\right\rangle _{A\left( t\right) }.
\end{equation}

The covariance of a signal might be thought of as the "average time"
multiplied by the instantaneous frequency or $\left\langle
t\vartheta^{\prime }\left( t\right) \right\rangle _{s}=\left\langle
t\vartheta^{\prime}\left( t\right) \right\rangle _{A\left( t\right) }.$When
time and frequency are uncorrelated with each other, then it is reasonable
to expect that $\left\langle t\vartheta^{\prime}\left( t\right)
\right\rangle =\left\langle t\right\rangle \left\langle \omega\right\rangle $%
, so the difference between the two is a measure of how time is correlated
to the instantaneous frequency. Thus, the covariance of the signal is 
\begin{equation}
Cov_{t\omega}=\left\langle t\vartheta^{\prime}\left( t\right) \right\rangle
-\left\langle t\right\rangle \left\langle \omega\right\rangle ,
\end{equation}
while the correlation coefficient, $r,$ is $r=\frac{Cov_{t\omega}}{\sigma
_{t}\sigma_{\omega}},$which is the normalized covariance. Real signals have
zero correlation coefficients as do signals of the form $A\left( t\right)
e^{i\omega_{0}t}$ or $S\left( \omega\right) =A\left( \omega\right)
e^{i\omega t_{0}}$, so signals with complicated phase modulation have a
non-zero correlation coefficient.

When dealing with more than one operator acting on a signal, we must be able
to interpret the action of multiple operators such as $\mathcal{\hat{A}\hat {%
B}}$ acting upon signals. Here $\mathcal{\hat{A}\hat{B}}$ is taken to mean $%
\mathcal{\hat{A}}$ acts on the signal followed by $\mathcal{\hat{B}}$ acting
on the signal. The commutator of $\mathcal{\hat{A}}$ and $\mathcal{\hat{B}}$
is 
\begin{equation}
\left[ \mathcal{\hat{A}},\mathcal{\hat{B}}\right] =\mathcal{\hat{A}\hat {B}-%
\hat{B}\hat{A}}\text{.}
\end{equation}
For example, the action of the time and frequency commutator on a signal is 
\begin{equation}
\left[ \mathcal{\hat{T}},\mathcal{\hat{W}}\right] s\left( t\right) =\left( 
\mathcal{\hat{T}\hat{W}-\hat{W}\hat{T}}\right) s\left( t\right) =is\left(
t\right) .
\end{equation}
This is analogous to the same result in quantum mechanics where the
commutator of the position and momentum operator is equal to $i$ when $%
\hbar=1$. The scale operator $\mathcal{\hat{C}}$ is defined as 
\begin{equation}
\mathcal{\hat{C}}=\frac{1}{2}\left[ \mathcal{\hat{T}},\mathcal{\hat{W}}%
\right] _{+}=\frac{1}{2}\left( t\frac{d}{dt}+\frac{d}{dt}t\right) .
\end{equation}
It can also be written as 
\begin{equation}
\mathcal{\hat{C}}=\mathcal{\hat{T}\hat{W}}+\frac{i}{2}.
\end{equation}
$\mathcal{\hat{C}}$ has the property that it transforms a signal $s\left(
t\right) $ according to 
\begin{equation}
e^{i\sigma\mathcal{\hat{C}}}s(t)=e^{\sigma/2}s(e^{\sigma/2}t)
\end{equation}
for a scaling parameter $\sigma$. Thus, the wideband ambiguity function can
be written as 
\begin{equation}
\chi_{WB}(\omega,\tau)=\sqrt{\alpha}\left\langle \,e^{-i\alpha\mathcal{\hat {%
C}}}e^{-i\tau\mathcal{\hat{W}}}\right\rangle _{s(t)},
\end{equation}
the expected value of the scaling and translation operators for a signal $%
s(t)e^{-it\pi f}$, which is equivalent to maximizing the signal to noise
ratio (SNR) at the receiver. We explore what physical interactions,
expressed in terms of operators, can be maximized.

\section{Physical Interactions}

While the primary scatterer produces the usual Doppler velocity and delay
which is equivalent to the range, the operator viewpoint may hold some
promise for finding interactions between the radar signal and the target
that extend beyond considerations of position and velocity related criteria.
Additional scatters can induce secondary characteristics into the return
signal, such as micro-Doppler, which can be incorporated into the design of
a receiver to maximize the possibility for detecting these types of
secondary target induced characteristics. In addition to a scalar signal,
higher dimensional waveform interactions can be considered as well, such as
how the polarization of materials affects the waveform. The cross ambiguity
function (CFA) symmetric form is defined as 
\begin{equation}
\chi_{r,s}(\omega,\tau)=\int_{-\infty}^{\infty}e^{-it\omega}\,q^{\ast}(t+%
\frac{\tau}{2})s(t-\frac{\tau}{2})\,dt,
\end{equation}
where $s(t)$ is the transmitted signal, while $q(t)$ is the correlation
signal and $\tau$ is the delay parameter. This is the traditional form for
the CFA. Instead of this form, a new type of CFA is proposed based on
quantum mechanics.

Any signal can be expressed as a complex vector. A new approach to signal
amplification is presented here \QTR{frametitle}{based on work by Aharonov
on amplification of the measurement of some operators in quantum phenomena 
\cite{Aharonov2005}. }Since any quantity that involves the usage of expected
values of complex signals can be expressed in the same mathematical form as
the quantum mechanical approach to signal amplification, the Aharonov
approach suggests a potential candidate for the \QTR{frametitle}{signal
amplification that is similar to a CFA. The classical equivalent to this is
what we choose to call cross correlation signal amplification. The
definition of the }cross correlation amplification of an observable $\hat{A}$
by the waveforms $\left\vert \Psi_{i}\right\rangle $ and $\left\vert
\Psi_{f}\right\rangle $ is: 
\begin{equation}
_{f}\left\langle \mathcal{\hat{A}}_{cross}\right\rangle _{i}=\frac {%
\left\langle \Psi_{f}|\hat{A}|\Psi_{i}\right\rangle }{\left\langle \Psi
_{f}|\Psi_{i}\right\rangle }
\end{equation}%
\QTR{frametitle}{\ where both }$\left\vert \Psi_{i}\right\rangle $%
\QTR{frametitle}{\ and }$\left\vert \Psi_{f}\right\rangle $\QTR{frametitle}{%
\ are normalized. Now, the obvious question is how does the cross
correlation measurement of an observable }$_{f}\left\langle \mathcal{\hat{A}}%
_{cross}\right\rangle _{i}$\QTR{frametitle}{\ differ from that of a normal
observable }$\hat{A}$\QTR{frametitle}{?}

Note that $\left\langle \Psi_{f}|\Psi_{i}\right\rangle \leq\left\langle
\Psi_{i}|\Psi_{i}\right\rangle \left\langle \Psi_{f}|\Psi_{f}\right\rangle =1
$, by the Cauchy-Schwartz inequality, so $\left\vert \left\langle \Psi
_{f}|\Psi_{i}\right\rangle \right\vert \leq1$. Thus, 
\[
\frac{1}{\left\vert \left\langle \Psi_{f}|\Psi_{i}\right\rangle \right\vert }%
\geq1, 
\]
and the effect of the denominator is to "magnify" the numerator provided
there is no counter balancing effect. Note that if $\hat{A}\left\vert \Psi
_{i}\right\rangle =\lambda_{\hat{A}}\left\vert \Psi_{i}\right\rangle $, so%
\[
_{f}\left\langle \mathcal{\hat{A}}_{cross}\right\rangle _{i}=\frac {%
\left\langle \Psi_{f}|\hat{A}|\Psi_{i}\right\rangle }{\left\langle \Psi
_{f}|\Psi_{i}\right\rangle }=\lambda_{\hat{A}}, 
\]
so there is no effect. When there is not this cancellation effect, there can
be a magnification, in some sense of the measurement of an operator. For an
electromagnetic wave, the operator interactions can be treated as either two
by two or four by four matrices. We consider only the two dimensional case.

\subsection{Multi-dimensional Interaction Operators}

Thus, the signal can be assumed to be of the form:%
\begin{equation}
\left\vert \Psi_{i}\left( t\right) \right\rangle =\left[ 
\begin{array}{c}
E_{1}^{i}\left( t\right) \\ 
E_{2}^{i}\left( t\right)%
\end{array}
\right] ,
\end{equation}
and the cross correlation signal is:%
\begin{equation}
\left\vert \Psi_{f}\left( t\right) \right\rangle =\left[ 
\begin{array}{c}
E_{1}^{f}\left( t\right) \\ 
E_{2}^{f}\left( t\right)%
\end{array}
\right] ,
\end{equation}
where the $E$'s can be real or complex. An interaction with a scattering
object can be thought as a matrix, $\hat{M}_{S}$, which acts on $\left\vert
\Psi_{i}\left( t\right) \right\rangle $ to give a return signal $\left\vert
\Psi_{R}\left( t\right) \right\rangle $, so 
\begin{equation}
\left\vert \Psi_{R}\left( t\right) \right\rangle =\hat{M}_{S}\left\vert
\Psi_{i}\left( t\right) \right\rangle .
\end{equation}
The cross correlation measurement amplification of operator $\hat{M}_{S}$ is 
\begin{equation}
_{f}\left\langle \mathcal{M}_{cross}\right\rangle _{i}\;\mathcal{=}\frac{%
\left\langle \Psi_{f}\left( t\right) |\Psi_{R}\left( t\right) \right\rangle 
}{\left\langle \Psi_{f}\left( t\right) |\Psi_{i}\left( t\right)
\right\rangle }.   \tag{weak}
\end{equation}
This example of amplification, which is analogous to spin systems in quantum
mechanics, applies to polarimetric radars. Consider the four polarization
matrices: 
\begin{equation}
\hat{\sigma}_{0}=\left[ 
\begin{array}{cc}
1 & 0 \\ 
0 & 1%
\end{array}
\right] ,\text{ \ \ }\hat{\sigma}_{1}=\left[ 
\begin{array}{cc}
1 & 0 \\ 
0 & -1%
\end{array}
\right] ,
\end{equation}%
\begin{equation}
\hat{\sigma}_{2}=\left[ 
\begin{array}{cc}
0 & 1 \\ 
1 & 0%
\end{array}
\right] ,\text{ }\hat{\sigma}_{3}=\left[ 
\begin{array}{cc}
0 & -i \\ 
i & 0%
\end{array}
\right] \text{.\ }
\end{equation}
The first operator, $\sigma_{0}$, acting on $\left\vert \Psi_{i}\left(
t\right) \right\rangle $ is the identity, so it is equivalent to the
previous no amplification case. Now, if the waveforms are normalized, $%
\left\vert E_{1}^{i}\left( t\right) \right\vert ^{2}+\left\vert
E_{2}^{i}\left( t\right) \right\vert ^{2}=1$ and $\left\vert E_{1}^{f}\left(
t\right) \right\vert ^{2}+\left\vert E_{2}^{f}\left( t\right) \right\vert
^{2}=1$, so%
\[
\tan\theta=\frac{E_{2}^{i}\left( t\right) }{E_{1}^{i}\left( t\right) }\text{
and }\tan\theta^{\prime}=\frac{E_{2}^{f}\left( t\right) }{E_{1}^{f}\left(
t\right) }, 
\]
thus, we have 
\[
\tan\theta\tan\theta^{\prime}=\frac{E_{2}^{i}\left( t\right) E_{2}^{f}\left(
t\right) }{E_{1}^{i}\left( t\right) E_{1}^{f}\left( t\right) }. 
\]
Note, that we treated amplitudes as real so the angles are real, this is not
necessary since complex angles are possible. The introduction of a complex
angle as well would introduce a second\ term which is imaginary that would
produce an additional effect on the imaginary component only. This
possibility will be discussed in a future paper.

Now,%
\[
\hat{\sigma}_{1}\left\vert \Psi_{i}\left( t\right) \right\rangle =\left[ 
\begin{array}{c}
E_{1}^{i}\left( t\right) \\ 
-E_{2}^{i}\left( t\right)%
\end{array}
\right] 
\]
so 
\begin{align}
\frac{\left\langle \Psi_{f}\left( t\right) \right\vert \hat{\sigma}%
_{1}\left\vert \Psi_{i}\left( t\right) \right\rangle }{\left\langle \Psi
_{f}\left( t\right) |\Psi_{i}\left( t\right) \right\rangle } & =\frac{%
E_{1}^{f}\left( t\right) E_{1}^{i}\left( t\right) -E_{2}^{i}\left( t\right)
E_{2}^{f}\left( t\right) }{E_{1}^{f}\left( t\right) E_{1}^{i}\left( t\right)
+E_{2}^{i}\left( t\right) E_{2}^{f}\left( t\right) }  \nonumber \\
& =\frac{1-\tan\theta\tan\theta^{\prime}}{1+\tan\theta\tan\theta^{\prime}}.
\end{align}
When $\theta\rightarrow-\frac{\pi}{4}$, 
\[
\frac{\left\langle \Psi_{f}\left( t\right) \right\vert \hat{\sigma}%
_{1}\left\vert \Psi_{i}\left( t\right) \right\rangle }{\left\langle \Psi
_{f}\left( t\right) |\Psi_{i}\left( t\right) \right\rangle }\rightarrow\frac{%
1+\tan\theta^{\prime}}{1-\tan\theta^{\prime}}\underset{\theta^{\prime}%
\rightarrow\frac{\pi}{4}}{\rightarrow}\infty, 
\]
so there can be amplification. In addition, we have 
\[
\hat{\sigma}_{2}\left\vert \Psi_{i}\left( t\right) \right\rangle =\left[ 
\begin{array}{c}
E_{2}^{i}\left( t\right) \\ 
E_{1}^{i}\left( t\right)%
\end{array}
\right] , 
\]
so%
\begin{align}
\frac{\left\langle \Psi_{f}\left( t\right) \right\vert \hat{\sigma}%
_{2}\left\vert \Psi_{i}\left( t\right) \right\rangle }{\left\langle \Psi
_{f}\left( t\right) |\Psi_{i}\left( t\right) \right\rangle } & =\frac{%
E_{1}^{f}\left( t\right) E_{2}^{i}\left( t\right) +E_{1}^{i}\left( t\right)
E_{2}^{f}\left( t\right) }{E_{1}^{f}\left( t\right) E_{1}^{i}\left( t\right)
+E_{2}^{i}\left( t\right) E_{2}^{f}\left( t\right) }  \nonumber \\
& =\frac{\tan\theta^{\prime}+\tan\theta}{\left( \tan\theta\tan\theta
^{\prime}+1\right) }  \nonumber \\
& =\frac{\sin\left( \theta+\theta^{\prime}\right) }{\cos\left(
\theta-\theta^{\prime}\right) }.
\end{align}
When $\theta-\theta^{\prime}\rightarrow\frac{\pi}{2}$, 
\[
\frac{\left\langle \Psi_{f}\left( t\right) \right\vert \hat{\sigma}%
_{2}\left\vert \Psi_{i}\left( t\right) \right\rangle }{\left\langle \Psi
_{f}\left( t\right) |\Psi_{i}\left( t\right) \right\rangle }=\underset{%
\varepsilon\rightarrow0}{\lim}\frac{\sin\left( 2\theta^{\prime }+\varepsilon+%
\frac{\pi}{2}\right) }{\cos\left( \varepsilon+\frac{\pi}{2}\right) }%
\rightarrow\infty, 
\]
so amplification is possible. Finally, we have 
\[
\hat{\sigma}_{3}\left\vert \Psi_{i}\left( t\right) \right\rangle =i\left[ 
\begin{array}{c}
-E_{2}^{i}\left( t\right) \\ 
E_{1}^{i}\left( t\right)%
\end{array}
\right] , 
\]
so%
\begin{align}
\frac{\left\langle \Psi_{f}\left( t\right) \right\vert \hat{\sigma}%
_{3}\left\vert \Psi_{i}\left( t\right) \right\rangle }{\left\langle \Psi
_{f}\left( t\right) |\Psi_{i}\left( t\right) \right\rangle } & =i\frac{%
-E_{1}^{f}\left( t\right) E_{2}^{i}\left( t\right) +E_{1}^{i}\left( t\right)
E_{2}^{f}\left( t\right) }{E_{1}^{f}\left( t\right) E_{1}^{i}\left( t\right)
+E_{2}^{i}\left( t\right) E_{2}^{f}\left( t\right) }  \nonumber \\
& =i\tan\left( \theta-\theta^{\prime}\right) ,
\end{align}
by using the trigonometric identity 
\[
\tan\left( \alpha\pm\beta\right) =\frac{\tan\alpha\pm\tan\beta}{1\mp
\tan\alpha\tan\beta}. 
\]
So amplification occurs as $\left( \theta-\theta^{\prime}\right) \rightarrow%
\frac{\pi}{2}$. Thus, the non-trivial operators $\hat{\sigma}_{1},\hat{\sigma%
}_{2},\hat{\sigma}_{3}$ can be amplified under the right conditions for the
components of cross-selection waveforms.

There are four additional operators to consider:%
\[
\hat{P}_{11}=\left[ 
\begin{array}{cc}
1 & 0 \\ 
0 & 0%
\end{array}
\right] , 
\]%
\[
\hat{P}_{12}=\left[ 
\begin{array}{cc}
0 & 1 \\ 
0 & 0%
\end{array}
\right] , 
\]%
\[
\hat{P}_{21}=\left[ 
\begin{array}{cc}
0 & 0 \\ 
1 & 0%
\end{array}
\right] , 
\]
and%
\[
\hat{P}_{22}=\left[ 
\begin{array}{cc}
0 & 0 \\ 
0 & 1%
\end{array}
\right] . 
\]
Now,%
\[
\hat{P}_{11}\left\vert \Psi_{i}\left( t\right) \right\rangle =\left[ 
\begin{array}{c}
E_{1}^{i}\left( t\right) \\ 
0%
\end{array}
\right] 
\]
so 
\begin{align}
\frac{\left\langle \Psi_{f}\left( t\right) \right\vert \hat{P}%
_{11}\left\vert \Psi_{i}\left( t\right) \right\rangle }{\left\langle
\Psi_{f}\left( t\right) |\Psi_{i}\left( t\right) \right\rangle } & =\frac{%
E_{1}^{f}\left( t\right) E_{1}^{i}\left( t\right) }{E_{1}^{f}\left( t\right)
E_{1}^{i}\left( t\right) +E_{2}^{i}\left( t\right) E_{2}^{f}\left( t\right) }
\nonumber \\
& =\frac{1}{\tan\theta\tan\theta^{\prime}+1}.
\end{align}
Note, the denominator goes to zero as $\tan\theta\tan\theta^{\prime
}\rightarrow-1$, while the numerator remains finite, so amplification is
possible for this operator. Also, if $\hat{P}_{11}$ is replaced by a
constant $a\hat{P}_{11}$, the amplification effect works as well. Now,%
\[
\hat{P}_{12}\left\vert \Psi_{i}\left( t\right) \right\rangle =\left[ 
\begin{array}{c}
E_{2}^{i}\left( t\right) \\ 
0%
\end{array}
\right] 
\]
so 
\begin{align}
\frac{\left\langle \Psi_{f}\left( t\right) \right\vert \hat{P}%
_{12}\left\vert \Psi_{i}\left( t\right) \right\rangle }{\left\langle
\Psi_{f}\left( t\right) |\Psi_{i}\left( t\right) \right\rangle } & =\frac{%
E_{1}^{f}\left( t\right) E_{2}^{i}\left( t\right) }{E_{1}^{f}\left( t\right)
E_{1}^{i}\left( t\right) +E_{2}^{i}\left( t\right) E_{2}^{f}\left( t\right) }
\nonumber \\
& =\frac{\tan\theta}{\tan\theta\tan\theta^{\prime}+1}
\end{align}
Note, the denominator goes to zero as $\tan\theta\tan\theta^{\prime
}\rightarrow-1$, while the numerator remains finite, so amplification is
possible for this operator. Also, if $\hat{P}_{12}$ is replaced by a
constant $a\hat{P}_{12}$, the amplification effect works as well. Now,%
\[
\hat{P}_{21}\left\vert \Psi_{i}\left( t\right) \right\rangle =\left[ 
\begin{array}{c}
0 \\ 
E_{1}^{i}\left( t\right)%
\end{array}
\right] 
\]
so 
\begin{align}
\frac{\left\langle \Psi_{f}\left( t\right) \right\vert \hat{P}%
_{21}\left\vert \Psi_{i}\left( t\right) \right\rangle }{\left\langle
\Psi_{f}\left( t\right) |\Psi_{i}\left( t\right) \right\rangle } & =\frac{%
E_{2}^{f}\left( t\right) E_{1}^{i}\left( t\right) }{E_{1}^{f}\left( t\right)
E_{1}^{i}\left( t\right) +E_{2}^{i}\left( t\right) E_{2}^{f}\left( t\right) }
\nonumber \\
& =\frac{\tan\theta^{\prime}}{\tan\theta\tan\theta^{\prime}+1}.
\end{align}
Note, the denominator goes to zero as $\tan\theta\tan\theta^{\prime
}\rightarrow-1$, while the numerator remains finite, so amplification is
possible for this operator. Also, if $\hat{P}_{21}$ is replaced by a
constant $a\hat{P}_{21}$, the amplification effect works as well. Now,%
\[
\hat{P}_{22}\left\vert \Psi_{i}\left( t\right) \right\rangle =\left[ 
\begin{array}{c}
0 \\ 
E_{2}^{i}\left( t\right)%
\end{array}
\right] 
\]
so 
\begin{align}
\frac{\left\langle \Psi_{f}\left( t\right) \right\vert \hat{P}%
_{22}\left\vert \Psi_{i}\left( t\right) \right\rangle }{\left\langle
\Psi_{f}\left( t\right) |\Psi_{i}\left( t\right) \right\rangle } & =\frac{%
E_{2}^{f}\left( t\right) E_{2}^{i}\left( t\right) }{E_{1}^{f}\left( t\right)
E_{1}^{i}\left( t\right) +E_{2}^{i}\left( t\right) E_{2}^{f}\left( t\right) }
\nonumber \\
& =\frac{\tan\theta\tan\theta^{\prime}}{\tan\theta\tan\theta^{\prime}+1}.
\end{align}
Note, the denominator goes to zero as $\tan\theta\tan\theta^{\prime
}\rightarrow-1$, while the numerator remains finite, so amplification is
possible for this operator. Also, if $\hat{P}_{22}$ is replaced by $a\hat {P}%
_{22}$, the amplification effect works as well.

Note, we have provided the necessary conditions under which these operators
can be amplified, but they are not sufficient. Sufficiency comes when
waveforms can be shown to obey the conditions the angles obey to produce
amplification. These conditions must be shown to be satisfied by specific
waveforms or classes of waveforms. In addition, noise has to be brought into
the mix.

\subsection{Scattering Operators}

The scattering operators for five specific structures are examined from the
viewpoint of amplification of operators. These scattering operators special
cases, two dimensional matrices, of the more general operators, four
dimensional matrices, found in Collett\cite{Collett2009}.

\begin{enumerate}
\item For a sphere, a plane, or triangular corner reflector oriented
horizontally, the scattering matrix is:%
\begin{equation}
S\left( h,r\right) =\left[ 
\begin{array}{cc}
1 & 0 \\ 
0 & 1%
\end{array}
\right] =\hat{\sigma}_{0}.
\end{equation}
Since this is the identity, there is no amplification effect. For a sphere,
a plane, or triangular corner reflector vertically polarized, the scattering
matrix is:%
\begin{equation}
\hat{S}\left( v,r\right) =\left[ 
\begin{array}{cc}
0 & i \\ 
i & 0%
\end{array}
\right] =i\hat{\sigma}_{2},
\end{equation}
so it can be amplified. (Note $h$ stands for horizontal polarization and $v$
stands for vertical polarization.)

\item For a dipole oriented along the vertical axis is :%
\begin{align}
\hat{S}\left( h,r\right) & =\left[ 
\begin{array}{cc}
1 & 0 \\ 
0 & 0%
\end{array}
\right] =\hat{P}_{11}, \\
& \text{and}  \nonumber \\
\hat{S}\left( v,r\right) & =\frac{1}{2}\left[ 
\begin{array}{cc}
1 & -i \\ 
-i & 1%
\end{array}
\right] =-\frac{i}{2}\hat{\sigma}_{2}+\frac{1}{2}\hat{\sigma}_{0}.
\end{align}
$\hat{S}\left( h,r\right) $ can be amplified, while the first term of $\hat{S%
}\left( v,r\right) $ can be amplified.

\item For a dipole oriented at the angle $\alpha$ from the positive
horizontal axis:%
\begin{align}
\hat{S}\left( h,r\right) & =\left[ 
\begin{array}{cc}
\cos^{2}\alpha & \frac{1}{2}\sin2\alpha \\ 
\frac{1}{2}\sin2\alpha & \sin^{2}\alpha%
\end{array}
\right] =\frac{1}{2}\sin2\alpha\hat{\sigma}_{2}+\cos^{2}\alpha\hat{P}%
_{11}+\sin^{2}\alpha\hat{P}_{22}, \\
& \text{and}  \nonumber \\
\hat{S}\left( v,r\right) & =\frac{1}{2}\left[ 
\begin{array}{cc}
e^{i2\alpha} & -i \\ 
-i & e^{-i2\alpha}%
\end{array}
\right] =\frac{-i}{2}\hat{\sigma}_{2}+e^{i2\alpha}\hat{P}_{11}+e^{-i2\alpha }%
\hat{P}_{22}.
\end{align}
Since $\hat{\sigma}_{2}$ and $\alpha\hat{P}_{11}+\beta\hat{P}_{22}\neq \hat{%
\sigma}_{0}$ can be individually amplified, then%
\begin{equation}
\alpha\hat{P}_{11}+\beta\hat{P}_{22}=\Xi=\left[ 
\begin{array}{cc}
\alpha & 0 \\ 
0 & \beta%
\end{array}
\right]
\end{equation}
where $\Xi=\Xi^{\dag}$which implies that $\alpha\hat{P}_{11}+\beta\hat{P}%
_{22}$ is Hermitian.

\item For a dihedral corner reflector oriented along the horizontal axis:%
\begin{align}
\hat{S}\left( h,r\right) & =\left[ 
\begin{array}{cc}
1 & 0 \\ 
0 & -1%
\end{array}
\right] =\hat{\sigma}_{1}, \\
& \text{and}  \nonumber \\
\hat{S}\left( v,r\right) & =\left[ 
\begin{array}{cc}
1 & 0 \\ 
0 & 1%
\end{array}
\right] =\hat{\sigma}_{0}.
\end{align}
The first operator can be amplified and the second can't.

\item For a right helix oriented at an angle $\alpha$ from the positive
horizontal axis:%
\begin{align}
\hat{S}\left( h,r\right) & =\frac{e^{-i2\alpha}}{2}\left[ 
\begin{array}{cc}
1 & -i \\ 
-i & 1%
\end{array}
\right] =\frac{e^{-i2\alpha}}{2}\left[ \hat{\sigma}_{0}-i\hat{\sigma}_{2}%
\right] , \\
& \text{and}  \nonumber \\
\hat{S}\left( v,r\right) & =\left[ 
\begin{array}{cc}
0 & 0 \\ 
0 & e^{-i2\alpha}%
\end{array}
\right] =e^{-i2\alpha}\hat{P}_{22}.
\end{align}
Clearly the matrix $\hat{S}\left( v,r\right) $ is amplified. For $\hat {S}%
\left( h,r\right) $, although $\hat{\sigma}_{0}$ is not amplified, the
component $\frac{e^{-i2\alpha}}{2}\left[ i\hat{\sigma}_{2}\right] $ is
amplified relative to it.

\item For a left helix oriented at an angle $\alpha$ from the positive
horizontal axis:%
\begin{equation}
\hat{S}\left( h,v\right) =\frac{e^{-i2\alpha}}{2}\left[ 
\begin{array}{cc}
1 & i \\ 
i & -1%
\end{array}
\right] =\frac{e^{-i2\alpha}}{2}\hat{\sigma}_{1}+\frac{ie^{-i2\alpha}}{2}%
\hat{\sigma}_{2}.
\end{equation}
Clearly this operator can be amplified.
\end{enumerate}

\subsection{Single Dimensional Interactions with Signals}

The goal of receiver design is to maximize the response of a receiver with
respect to the return signal $s_{R}\left( t\right) $. The functional form is 
$s_{R}\left( t\right) =s(at+\tau)$ where $\tau$ is the (delay) time it takes
the signal to reach the target and return to the receiver, and $a$ is
dilation of the time axis due to the motion of the object. This is
accomplished by taking the inner product of $s_{R}\left( t\right) $ with $%
s^{\ast}(t)$ and integrating, so we are computing the Fourier transform of
the product $\,s^{\ast}(t)s(at+\tau)$:%
\begin{equation}
\frac{A}{\sqrt{\pi}}\left\langle
\,s(t)e^{it\omega},s(at\pm\tau)\right\rangle =\frac{A^{\prime}}{\sqrt{\pi}}%
\left\langle \,e^{ia\mathcal{\hat{C}}}e^{i\tau\mathcal{\hat{W}}%
}\right\rangle ,
\end{equation}
which is the expected value of the operators for scale $e^{ia\mathcal{\hat{C}%
}}$ and the operator for time shift $e^{i\tau\mathcal{\hat{W}}}$. Trying to
maximize the reception SNR has led to the ambiguity function which can be
interpreted as the expected value of two specific operators for a given
signal $s(t)$.

The \QTR{frametitle}{non-uniform Doppler effect can be used to illustrate
this operator viewpoint. The effect of non-uniform Doppler on the radar
waveform can be determined by }the application of the relativistic boundary
conditions to the D'Alembert solution to the wave equation\QTR{frametitle}{%
\cite{Gray2003}}. The scattered waveform in terms of the incident waveform
becomes 
\begin{equation}
g(\tau)\simeq f\left( \tau-\frac{2r(\tau)}{c}\right) .
\end{equation}
For a dynamic system characterized by single parameter $\alpha$, then a
dynamic variable $u$ evolves along a path in configuration space. The
configuration of the system describes a curve along $\alpha$. Consider the
commutator equation \ 
\begin{equation}
\frac{du}{d\alpha}=\left[ u,\hat{G}\right] .
\end{equation}
Here, $\hat{G}$ generates the trajectory $u=u(\alpha)$ and $\alpha$ can be
viewed as geometrical parameter. Expanding $u(\alpha)$ in a Taylor series
yields\cite{Souza1990} a Taylor series, thus the generator equation can be
used to replace the dynamics with the operator equation\cite{Jordan2004} 
\begin{equation}
u(\alpha)=u_{0}+\alpha\left. \left[ u,\hat{G}\right] \right\vert _{\alpha=0}+%
\frac{\alpha^{2}}{2!}\left. \left[ \left[ u,\hat{G}\right] ,\hat{G}\right]
\right\vert _{\alpha=0}+...\text{ }=\exp\left( \alpha\hat {G}\right) \left.
u(\alpha)\right\vert _{\alpha=0}\text{.}
\end{equation}
For physical systems, it is evident that the generator of dynamics is time,
so any function of time can be thought of as being generated by an operator, 
$\hat{G}$, acting on $u\left( t\right) $, so it can be thought of being
"generated" by that operator. It is evident how to "generate" any function
of a parameter using operator methods\cite{Fernandez2002}. For a given $%
r(\tau)$, we can assume it is generated by a equation such as $%
r(\tau)=\exp\left( k\hat{G}\right) \left. x(\tau)\right\vert _{\tau=0}$, so 
\begin{equation}
f\left( \tau-\frac{2r(\tau)}{c}\right) =f(\tau-\alpha\exp\left( \hat {G}%
\right) \left. r(\tau)\right\vert _{\tau=0})=\exp\left( \alpha \mathcal{\hat{%
H}}\right) s\left( \tau\right) ,
\end{equation}
where $\mathcal{\hat{H}}$ depends on the specifics of the interaction. For
example, $\mathcal{\hat{H}}$ would be a comb operator in the frequency
domain for a periodic function. In this case, we are estimating the expected
value $\left\langle \exp\left( \tau\mathcal{\hat{H}}\right) \right\rangle $
at the receiver. Since any scalar interaction on the waveform can be thought
of as the action of an operator on the broadcast waveform, a more general
ambiguity function can always be defined as 
\begin{equation}
\chi_{\hat{O}}(\omega,\tau)=\frac{A}{\sqrt{\pi}}\int_{-\infty}^{\infty
}e^{-it\omega}\,s^{\ast}(t)\exp\left( \tau\mathcal{\hat{H}}\right)
s(t)dt=\left\langle e^{-it\omega}\,s^{\ast}(t),\exp\left( \tau\mathcal{\hat {%
H}}\right) s(t)\right\rangle .
\end{equation}

For the remainder of the discussion, we assume the signal is not normalized.
The typical signal processing application is to minimize the effect of the
noise $\tilde{n}$ so as to maximize the signal-to-noise ratio (SNR) for a
received signal $\tilde{y}$. In order to understand how to do this, one uses
a linear model for the combination of signal plus noise $\tilde{y}=s\left(
t\right) +\tilde{n}$. The response to an input $f(t)$ of a system function $%
h(t)$ is the response $g(t)$, which at a time $t_{0}$ is%
\[
g(t_{0})=\int_{-\infty}^{\infty}s(t)h(t-t_{0})dt=\frac{1}{2\pi}\int_{-\infty
}^{\infty}S\left( \omega\right) e^{i\omega t_{0}}H\left( \omega\right)
d\omega. 
\]
Here we wish to determine the maximum value of $g(t_{0})$; this allows us to
maximize SNR depending on which of several integral constraints that are
specific to the problem being considered. The SNR depends on the mean
squared constraint under consideration: it could be based on the energy
spectrum $\left\vert S\left( \omega\right) \right\vert ^{2}$, it could be
based on the constrained energy spectrum $\left\vert S\left( \omega\right)
\right\vert ^{2}\left\vert R(\omega)\right\vert ^{2}$, it could be based on
multiple constraints such as higher order moments of the energy spectrum, or
it could be based on amplitude constraints. Each constraint leads to a
different choice for the system response function $H\left( \omega\right) $.

If we have a specified energy 
\begin{equation}
E=\left\langle s(t),s(t)\right\rangle =\frac{1}{2\pi }\int_{-\infty
}^{\infty }\left\vert S\left( \omega \right) \right\vert ^{2}d\omega ,
\end{equation}%
then by the Cauchy-Schwartz inequality 
\begin{align}
\left\vert \int_{-\infty }^{\infty }S\left( \omega \right) e^{i\omega
t_{0}}H\left( \omega \right) d\omega \right\vert ^{2}& \leq \int_{-\infty
}^{\infty }\left\vert S\left( \omega \right) \right\vert ^{2}d\omega
\int_{-\infty }^{\infty }\left\vert e^{i\omega t_{0}}H\left( \omega \right)
\right\vert ^{2}d\omega   \nonumber \\
& \leq \frac{E}{2\pi }\int_{-\infty }^{\infty }\left\vert e^{i\omega
t_{0}}H\left( \omega \right) \right\vert ^{2}d\omega .
\end{align}%
The inequality becomes an equality only if 
\begin{equation}
S\left( \omega \right) =ke^{i\omega t_{0}}H^{\ast }\left( \omega \right) ,
\end{equation}%
so the maximum value for $g(t_{0})$ is obtained by the choice 
\begin{equation}
s(t)=kh^{\ast }(t_{0}-t)
\end{equation}%
since $H^{\ast }\left( \omega \right) \longleftrightarrow h^{\ast }(-t)$ and 
$k$ is an arbitrary constant.\QTR{frametitle}{\ }For a linear system $\tilde{%
y}\left( t\right) =s\left( t\right) +\tilde{n}\left( t\right) $ with an
impulse response $h(t)$, the output $\widehat{u}(t)$ is 
\begin{equation}
\tilde{u}(t)=\tilde{y}\left( t\right) \ast h(t)=\tilde{u}_{s}(t)+\tilde{u}%
_{n}(t)
\end{equation}%
where $\tilde{u}_{s}(t)=s\left( t\right) \ast h(t)$ and $\tilde{u}_{n}(t)=%
\tilde{u}\left( t\right) \ast h(t)$. Now the response to the signal $s\left(
t\right) $ is 
\begin{equation}
\tilde{u}_{s}(t_{0})=\frac{1}{2\pi }\int_{-\infty }^{\infty }S\left( \omega
\right) e^{i\omega t_{0}}H\left( \omega \right) d\omega ,
\end{equation}%
so 
\begin{equation}
\left\vert \tilde{u}_{s}(t_{0})\right\vert ^{2}\leq \int_{-\infty }^{\infty
}S\left( \omega \right) e^{i\omega t_{0}}H\left( \omega \right) d\omega
\int_{-\infty }^{\infty }S\left( \omega \right) e^{i\omega t_{0}}H\left(
\omega \right) d\omega ;
\end{equation}%
this is what Papoulis has called the \emph{Matched Filter Principle}\cite%
{Papoulis1977}. From the operator perspective, the operator acting on the
signal should replace the operator acting on the system response function in
this argument, so 
\[
g(t_{0})=\int_{-\infty }^{\infty }s\left( \tau \right) \exp \left( \alpha 
\mathcal{\hat{H}}\right) h(\tau -\tau _{0})d\tau =\frac{1}{2\pi }%
\int_{-\infty }^{\infty }S\left( \omega \right) e^{i\omega \tau _{0}}R\left(
\omega \right) d\omega .
\]%
where 
\begin{equation}
\left\langle e^{i\omega \tau },\exp \left( \alpha \mathcal{\hat{H}}\right)
h(\tau -\tau _{0})\right\rangle =\left\langle e^{i\omega \tau },\exp \left(
\alpha \mathcal{\hat{H}}\right) e^{\tau _{0}\frac{d}{d\tau }}h(\tau
)\right\rangle =e^{i\omega \tau _{0}}\left\langle e^{i\omega \tau },\exp
\left( \alpha \mathcal{\hat{H}}\right) h(\tau )\right\rangle =e^{i\omega
\tau _{0}}R\left( \omega \right) 
\end{equation}%
since the operators commute. When 
\begin{equation}
\exp \left( \alpha \mathcal{\hat{H}}\right) =\,e^{ia\mathcal{\hat{C}}%
}e^{i\tau \mathcal{\hat{W}}},
\end{equation}%
and the optimum choice is a rescaled version of the transmitted signal time
scale $t_{0}\rightarrow at\pm \tau $, the wideband matched filter.

The Matched Filter Principle is quite general and can be used to introduce a
variety of constraints, which are equivalent to a cost function minimization
approach. For example, if one wanted to maximize the response to the
derivative of the energy $E_{1}$, while requiring the energy to be
normalized, then one has 
\begin{equation}
E_{1}=\int_{-\infty}^{\infty}\left\vert s^{\prime}\left( \tau\right)
\right\vert ^{2}d\tau=\frac{1}{2\pi}\int_{-\infty}^{\infty}\omega
^{2}\left\vert F\left( \omega\right) \right\vert ^{2}d\omega
\end{equation}
so $\left\vert s^{\prime}\left( \tau\right) \right\vert \leq\sqrt[4]{E_{1}}$
with equality at time $t_{0}$ if $s\left( \tau\right) =\sqrt[4]{E_{1}}%
\exp\left( \frac{\tau-\tau_{0}}{\sqrt{E_{1}}}\right) $. In general, using
this approach, arbitrary constraints can be considered. If we have a signal $%
\exp\left( \alpha\mathcal{\hat{H}}\right) s\left( \tau\right) $ of where the
energy of $s\left( \tau\right) $ is $E$, that we want maximize the system
response $g\left( t_{0}\right) $ of the system $h\left( t\right) $, then to
obtain the maximum subject to the constraints 
\begin{equation}
\int_{-\infty}^{\infty}s\left( t\right) \Phi_{i}\left( t\right)
dt=\vartheta_{i},
\end{equation}
where the functions $\Phi_{i}\left( t\right) $ and constraints $\vartheta_{i}
$ are given. Then, with the definition%
\begin{equation}
u_{i}\left( t\right) =\Phi_{i}\left( t\right) -\frac{\vartheta_{i}}{S(0)},
\end{equation}
that the constraint equation becomes:%
\begin{equation}
\int_{-\infty}^{\infty}s\left( t\right) u_{i}\left( t\right) dt=0
\end{equation}
because $S(0)$ is the area of $s\left( t\right) $. Thus, it follows that the
system response is 
\begin{equation}
g(\tau_{0})=\int_{-\infty}^{\infty}s\left( \tau\right) \left[ \exp\left(
\alpha\mathcal{\hat{H}}\right) e^{\tau_{0}\frac{d}{d\tau}}h(\tau
)+\sum\limits_{i=1}^{n}\beta_{i}u_{i}\left( \tau\right) \right] d\tau,
\end{equation}
for arbitrary $\beta_{i}$. Therefore, $g(\tau_{0})$ can be bounded by 
\begin{equation}
\left\vert g(\tau_{0})\right\vert ^{2}\leq
E\int_{-\infty}^{\infty}\left\vert \left[ \exp\left( \alpha\mathcal{\hat{H}}%
\right) e^{\tau_{0}\frac{d}{d\tau }}h(\tau)+\sum\limits_{i=1}^{n}%
\beta_{i}u_{i}\left( \tau\right) \right] \right\vert ^{2}d\tau.
\end{equation}
Equality is achieved if 
\begin{equation}
s\left( \tau\right) =\left[ \exp\left( \alpha\mathcal{\hat{H}}\right)
e^{\tau_{0}\frac{d}{d\tau}}h^{\ast}(\tau)+\sum\limits_{i=1}^{n}\beta_{i}^{%
\ast}u_{i}^{\ast}\left( \tau\right) \right] .
\end{equation}
This gives a method for choosing the correlation waveform to achieve maximum
response for a given set of constraints.

\section{Conclusions}

The operator method is a much richer way to look at the radar measurement
problem because of its ability to produce a wide variety of distributions
associated with the information contained in a signal. In particular, it is
possible to put the ambiguity function in a wider context as part of a
general theory of measurement. There is a much greater freedom of
description of the same physical situation which suggests that we can find
information present in waveforms that a waveform designer would not think to
look for. This approach to incorporating quantum mechanical ideas has been
championed by Baraniuk\cite{Baraniuk1995}\cite{Baraniuk1998} recently by
extending the Hermitian operator approach in quantum mechanics to unitary
operators in signal processing. The specifics of the type of operators
matter relative to the physics of the interaction of the target with the
waveform, so this may be important for future extensions of this work.

\textbf{Acknowledgement:} This work was supported by NSWCDD In-House
Laboratory Independent Research (ILIR) Program.

\end{document}